White paper on

# Quantum Information Science and Technology for Nuclear Physics

Input into U.S. Long-Range Planning, 2023

## Table of Contents

# Executive Summary

The Department of Energy (DOE), the National Science Foundation (NSF), the National Institute of Standards and Technology (NIST), and other agencies are investing substantially in basic research in quantum information science and technology (QIST) and its applications, and this funding is anticipated to continue beyond the next decade. During the previous long-range planning (LRP) period, research at the interface of nuclear physics (NP) and QIST has benefited substantially from this increased investment. It has resulted in important advances and benchmarking for future research and in growing interdisciplinary collaborations.

Advances made over the past two decades in atomic, molecular, optical, and material sciences, as well as cryogenic infrastructure, are accelerating the development of quantum sensors and quantum integrated systems, and in some cases are providing revolutionary approaches to historically inaccessible problems. Quantum sensors are already in use in certain high-priority NP programs, such as neutrinoless double-$\beta$ decay, neutrino mass measurements, sterile-neutrino searches, precision tests of fundamental symmetries, permanent electric dipole moment searches, and as probes of rare and exotic processes. Their targeted use in NP continues to grow, and expanding research and development in this area, including through investments in facilities at national laboratories and universities, is essential.

With a substantial increase in computing capability enabled by superposition and entanglement, quantum computing and quantum simulation have the potential to provide capabilities that are unique, and that far exceed those possible with classical computation alone in addressing specific challenges in NP. These range from properties of nuclei and dense matter to dynamics of non-equilibrium strongly-interacting matter or neutrino gasses. Furthermore, quantum-information tools are beginning to guide the design of more efficient classical NP simulations, and quantum entanglement is now being investigated as a new guiding principle in our understanding of NP phenomena and the Standard Model. The community will substantially benefit from strengthening its efforts in the co-design of quantum-simulation devices for the NP program, better engaging with the DOE Quantum Testbed Program, and further developing reliable access to forefront quantum hardware, including industry platforms and other testbeds at national laboratories and universities. Programs and partnerships that enable collaborations across NP in QIST are increasingly valuable.

The NP community's understanding of the interactions between particles and matter serves as an asset in the development of quantum computers and quantum sensors, and of a sustainable quantum workforce. For example, nuclear physicists' expertise in shielding against cosmic rays and in the development of radio-pure materials for rare-event searches, is playing an important role in increasing the coherence times of next-generation qubits for a range of computing and sensing applications. Likewise, the co-design of physics-informed hardware and hardware-aware algorithms for NP systems is leading to new and improved capabilities and efficiencies for quantum simulations in NP and other domain sciences. We encourage strengthening such cross-disciplinary activities and collaborations between the NP and QIST communities.

A diverse and sustainable quantum-ready workforce is a necessity for both the NP and QIST communities. Owing to its broad application space, QIST attracts science talent into NP from a variety of backgrounds. This talent is then empowered by acquiring skills at the nexus of emerging quantum technology and computing trends in NP research. Recruiting and training this new generation of researchers will accelerate the development and integration of quantum technologies in NP research, and propel growth and diversity in the NP workforce.

The *Quantum Information Science for US Nuclear Physics Long Range Planning–2023* workshop, therefore, unanimously endorses the following two recommendations.

**Solving grand challenges in Nuclear Physics (NP) requires the development of fundamentally new tools for simulation and sensing. U.S. Nuclear Science is in an early stage of benefiting from and contributing to the advancement of Quantum Information Science and Technology (QIST). To capitalize on this progress, we recommend:**

- **Increasing support for driving advances at the interface of NP and QIST to uniquely address Nuclear Science objectives.**

    This support will advance the development of quantum sensor technology for NP; enhance the (co-)development, integration, and application of quantum-based simulation and computation hardware and techniques for NP; grow cross-cutting research and partnerships that leverage NP expertise to accelerate advances in QIST (including access to forefront hardware and fabrication); and expand the training of, and robust professional pathways for, a diverse and inclusive quantum-ready workforce for NP, with cross-disciplinary collaborations in QIST.

- **Establishing an 'NP *Quantum Connection*' that will realize the transformational potential of QIST in addressing NP grand challenges.**

    This national initiative will enable a community-wide integration of quantum sensing and simulation in NP research; facilitate sharing of resources and expertise among NP, interagency programs, and the national and international QIST community; support bridge junior faculty and scientist positions, postdoctoral fellowships, and graduate and undergraduate students; and strengthen ties with the QIST community, technology companies, and other domain sciences.

# Introduction

Quantum information science and technology (QIST) uses the fundamental laws of nature to detect and acquire, manipulate, transmit and store information in ways that exceed the capabilities of classical physics. Entanglement and coherence, which are essentially quantum, are at the heart of these new capabilities. The range of possible applications of QIST is significant, including advancing nuclear-physics (NP) research and technology. The systems and environments relevant to NP research range from high-density, non-equilibrium strongly-interacting many-body systems, to technology that is sensitive to the most-weakly interacting particles. The use of entanglement and coherence in developing and understanding these systems is now at the forefront of NP research, development, and technology. Concurrently, rapid advances in QIST during the past two decades suggests that revolutionary advances in NP experiment, computing, and theory may be possible in the near future. Beyond understanding and utilizing coherence and entanglement in NP systems, quantum sensing and quantum simulation are two main areas in which mutually beneficial advances to NP and QIST are expected.

QIST was not considered in depth during the previous long-range planning activity, but emerged as being of significance to NP during the subsequent period. Since then, there have been a number of high-profile developments in QIST, both scientifically and organizationally. In particular, the National Quantum Initiative (NQI) Centers have been established, and the agencies are providing significant funding for QIST and related research within the domain sciences. Quantum computers have begun to be accessible to NP researchers via the cloud, through collaborations with technology companies, DOE testbeds, NQI centers, or collaborations with colleagues at universities and national laboratories. Collaborations have been established, through grassroots efforts or via funded proposals, that have brought together QIST and NP researchers to develop quantum sensors, improve qubits, conceive quantum algorithms, and co-design aspects of quantum computers.

A vision for the potential benefits of QIST for NP and, conversely, the potential impact of the NP knowledgebase on QIST, has been captured in earlier whitepapers [1,2] and the 2019 NSAC Subcommittee report on Quantum Information Science [3]. In preparation for the 2023 NSAC Long Range Plan (LRP), members of the Nuclear Science community gathered to discuss the current state of, and plans for further leveraging opportunities in, QIST in NP research at the *Quantum Information Science for U.S. Nuclear Physics Long Range Planning* workshop [4], held in Santa Fe, New Mexico on January 31—Feb 1, 2023. The workshop, jointly-sponsored by Los Alamos National Laboratory (LANL) and the InQubator for Quantum Simulation (IQuS), included 45 in-person participants and 53 remote attendees. The outcome of the workshop identified strategic plans and requirements for the next 5-10 years to advance quantum sensing and quantum simulations within NP, and to develop a diverse quantum-ready workforce. The plans include resolutions endorsed by the participants to address the compelling scientific opportunities at the intersections of NP and QIST. These endorsements are aligned with similar affirmations by the LRP Computational Nuclear Physics and AI/ML Workshop, the Nuclear Structure, Reactions, and Astrophysics LRP Town Hall, and the Fundamental Symmetries, Neutrons, and Neutrinos LRP Town Hall communities, see Appendix III.

While there has been substantial progress on integrating advances in QIST in NP as evidenced in the following, the efforts, both for sensing and for simulation, are often sub-critical. Part of the success of the whitepaper workshop has been to make the broader community aware of this progress. We propose to formalize and enhance sharing of developments and thereby better link efforts at different institutions through the proposed "NP Quantum Connection". This organization would be able to play a particularly important role in workforce development.

During breakout discussion sessions on quantum sensing and quantum computing and simulation, and with a focus on multidisciplinary connections and workforce development, workshop participants provided valuable input on four main questions:

- ➢ How will quantum sensing and simulation advance NP?
- ➢ What are the strategic goals to accelerate?
- ➢ What do we need in the next 5-10 years to achieve those goals?
- ➢ How does quantum sensing and simulation in NP impact other NSF and DOE areas?

This whitepaper summarizes the conclusions reached during these community discussions, and provides further context and explanation for the Recommendations.

## Quantum Sensing

Quantum sensors have many applications, including in imaging, communication, medicine, deep-space exploration, national security, and the domain sciences. Their integration into NP applications importantly advances NP research, and also accelerates the transfer of such quantum technologies into the supply chain. Further, expertise in the NP community helps drive forward the development of quantum sensors and other quantum devices.

### How will quantum sensing advance NP?

In many instances, quantum-enabled sensors are already being deployed in high priority nuclear physics initiatives. These technologies can enhance the reach of a myriad of different types of probes: from the measurement of non-zero neutrino masses, to searches for beyond Standard Model physics via light and heavy sterile neutrino states, coherent neutrino nuclear scattering, precision measurements of nuclear-$\beta$ decay and its energy decay spectrum, dark photons, and precision tests of the CKM matrix unitarity.

Currently deployed technologies involve instruments sensitive to low energy transitions, such as superconducting tunnel junctions (STJs) and transition-edge sensors (TESs). Looking forward, entanglement in many-body systems can be used as a tool to reduce fluctuations below the standard quantum limit (SQL). Such a reduction of fluctuations can be realized with so-called squeezed states. For atomic sensors, a significant boost to the sensitivity of the phase sensor is possible when using entangled states.

Superconducting nanowire particle detectors are being developed for NP applications, in particular, for the Electron Ion Collider (EIC) and for experiments at Jefferson Lab. Building on the existing superconducting nanowire single photon detector (SNSPDs) technology, SNSPDs can now operate in magnetic fields above 5 T, at high rates, and with nearly zero dark count rate. The detection occurs through the excitation of a pair of quasi-electrons which rapidly scatter, generating a concentration of quasi-particles high enough to suppress the superconducting state across the width of the superconductor. This hot-spot formation process proceeds rapidly and is one of the fastest and most sensitive means of detecting individual quantum excitations. Ongoing research aims to leverage these characteristics for a broader range of unconventional applications for SNSPDs from low-energy ion detectors to high-energy particle tracking.

Certain rare nuclei possess novel properties and are poised to benefit, and benefit from, quantum techniques in the future. These "nuclear sensors" have the potential to achieve unprecedented precision in searches for Beyond Standard Model (BSM) physics and metrology. Two concrete examples are searches for electric dipole moments (EDMs) in octupole-deformed nuclei and nuclear clocks based on low-energy isomer levels. Future improvements to measurement precision will likely employ quantum techniques such as squeezing, entanglement, or coherent molecule assembly to achieve the ultimate potential sensitivity to BSM physics.

Significant sensing tasks in NP rely on an array of sensors working collectively. Generating entangled states in distributed systems is, however, challenging. A novel approach uses variational quantum circuits to tailor the entangled states to a configuration that can maximize measurement sensitivity. Such circuits have been extensively explored in quantum machine learning as quantum neural networks for optimization and inference applications, but they have, to date, rarely been used in sensing settings. In NP, the adoption of variationally-enhanced, distributed quantum sensing could significantly boost the sensitivity of axion searches, and in measurements of the neutrino mass via $\beta$-decay using cyclotron radiation emission spectroscopy (CRES).

## What are the strategic priorities in the next 5-10 years?

To maximize the investments made in NP to tackle the grand challenge goals defined in the field with the added capabilities of quantum sensing, we recommend prioritizing the following activities:

*Leverage Existing Quantum 1.0 Technologies to Full Advantage.* Several existing technologies like transition edge sensors and superconducting tunnel junctions (informally referred to as "Quantum 1.0") are already in use in a number of NP experiments. However, there are likely many high impact applications that have not yet been realized. A high priority is to create an environment where the full application range of Quantum 1.0 devices can be explored. Among the many limiting factors that have prevented such exploration up to now is the short supply of such sensors. The capability to produce them exists at just a few institutions and private vendors, a difficulty that is closely connected to the limited quantum workforce. The resulting bottleneck limits individual research groups' ability to acquire devices for R&D. The democratization of Quantum 1.0 device access, e.g., by establishing more fabrication capability at national laboratories and universities (that can seed private enterprises) will be required to realize the full potential of Quantum 1.0 sensors for NP experiments. To be consistent with our goals of inclusivity, these new capabilities should be available to all in the US NP community.

*Coordinated Interagency Development of Quantum 2.0 Technologies.* "Quantum 2.0" sensors —i.e., sensors that take advantage of entanglement and/or coherence to enhance sensitivity to a particular phenomenon—will lead to breakthroughs in NP physics experiments. We have already seen the use of squeezed states to improve sensitivity in searches for axions and gravitational waves. Given the infancy of Quantum 2.0 within NP, broad, blue-sky ideas need to be supported. This development of Quantum 2.0 technologies would be greatly enhanced by cooperation with agencies where there is mutual interest—for example, HEP and the NNSA's Office of Defense Nuclear Nonproliferation (NA22) are obvious partners.

*Leverage NP expertise to help the QIST community go beyond Noisy Intermediate Scale Quantum (NISQ) technology that limits solid state devices.* The impact of ionizing radiation on superconducting qubit decoherence was first measured by a cross-disciplinary collaboration of NP researchers and QIST experts. The NP community should play a significant role in reducing the impact of ionizing radiation by employing NP techniques such as device design, radiation transport modeling, shielding, and low radioactivity materials sourcing and production. The NP community should seek more opportunities for collaboration with the QIST community where NP expertise can be brought to bear on improving quantum technology.

## What do we need in the next 5-10 years?

The next decade offers a unique opportunity to realize the priorities outlined above. Below are ways in which these priorities can be achieved.

*Facilitate Access to Resources, Technology and Expertise.* R&D to develop new quantum sensors or to effectively deploy sensors on experiments is restricted by the limited number of institutions and vendors able to produce these devices. NP should encourage commercialization and large-scale production of mature Quantum 1.0 sensor technologies (e.g., STJs, TESs, MKIDs) and associated readout hardware (e.g., SQUIDs, JPAs, TWPAs). Within the proposed *NP Quantum Connection initiative*, a foundry, or network of facilities, could be established to democratize development of improved Quantum 2.0 sensors, and provide expert assistance for device design to effectively utilize them.

*Continue to Address Basic Problems Through Dedicated R&D.* There are several known limitations of Quantum 1.0 sensors that can be addressed in the next few years. In general, faster sensor response times, higher channel counts, and more effective multiplexing (readout of many sensors on a single line) are needed. In most cases there is also a need to better understand the physics behind device operation. For example, the processes by which high energy (keV and above) radiation transfers its energy to broken Cooper pairs in a superconducting sensor is not understood in detail. This could be addressed by a program of modeling and experimentation. NP would benefit from a program specifically supporting detector development, such as HEP's Advance Detector R&D. Some promising quantum sensing technologies are not currently associated with any NP experiment, making them unnatural residents of NP's research ecosystem. A dedicated program of quantum sensor R&D would accelerate development and lead to new ideas on how to engage quantum sensors in the solutions of NP's core science missions.

*Secure the Rare Isotope Supply.* In some cases, atoms of rare isotopes are used as quantum sensors (and potentially quantum memories). Careful planning is required to maintain a stable and predictable supply of these isotopes, and partnership with the DOE Isotope Program is advised.

## How does QS in NP impact other NSF and DOE areas?

Advances outlined above in NP have potential benefits for other science programs undertaken in the US. Below we list a few of such cases.

*Nonproliferation Applications.* Precise isotopic assays of nuclear materials can provide information on material source and history. Obtaining this information via detected radiation, however, is challenging due to the close proximity and complexity of relevant lines, orders of magnitude variation in line intensity, and in cases, limited or incomplete nuclear data. Proof of principle quantum sensors based on x-ray and gamma-ray spectrometers have already demonstrated the capability to isolate weak gammas lines, such as those from $^{242}$Pu, that cannot be separated using conventional technology. For the first time, sensor arrays of this type allow non-destructive assay of safeguards-relevant samples without special preparation.

*Dark Matter.* Over the past 10 years the dark matter direct detection community has focused on low-mass fermionic particle candidates with masses much below the GeV scale. Due to the low relative velocity in the galactic dark matter halo, the energies imparted in detector targets is on the scale of meV. As such, there has been much interest in the dark matter community in ultra-low threshold precision detectors. This community would benefit greatly from the development of quantum sensors for NP applications. This is similar to the logic that motivates Quantum 1.0 sensors in BSM searches in very low-recoil energy detection for coherent neutrino-nucleus scattering.

*High-channel-count Multiplexing.* Cryogenic hardware used to read out quantum sensors, including in NP experiments, has largely converged on superconducting microwave resonators and/or SQUIDs. The need to read out large quantum sensor arrays has driven the development of novel SQUID/resonator readout architectures with progressively increasing multiplexing factors. These readout technologies may have potential uses outside of quantum sensing, e.g., the readout of future large quantum computing qubit arrays.

*Large Sensor Arrays for Astronomy.* High-pixel count, low-threshold sensor arrays with sensitivity to single THz photons are needed for next generation astrophysical instruments to perform high red-shift surveys to study stellar and galactic formation. Such detectors will rely on the same type of readout and multiplexing solutions that will be useful in NP applications as discussed above.

*Medical Physics.* Quantum sensing will improve several aspects of detection in medical physics. Nuclear data applications have been discussed above—high resolution, low-energy, low-threshold detectors can support data on isotopes that are promising for new medical diagnostics and therapies (e.g., cancer therapy with Auger-emitting isotopes). It has also been shown that quantum imaging can improve images obtained by positron emission tomography (PET) by taking advantage of the entanglement correlation between e+/e- annihilation photons to beat statistical noise limits.

## Sidebar: **Radiation and Quantum Computing**

In order to make quantum computing a viable and usable technology, the underlying units of computation, qubits, must simultaneously exhibit high fidelity and long coherence times. Over the past two decades, advances in device design, fabrication, and materials have increased coherence times by six orders of magnitude. Nonetheless, to realize the full promise of quantum computing, far longer coherence times will be needed to achieve the operational fidelity required for fault-tolerant computation. Coherence for superconducting qubits can be spoiled by an excess density of quasi-particles, and quasi-particle densities far exceed what is naively expected from thermal equilibrium. Recent measurements made by several groups have shown that one creeping contribution to quasi-particle poisoning appears due to ionizing radiation stemming from external gamma radiation, cosmic rays, and radiogenic contamination of materials surrounding the qubit. This source of quasi-particle poisoning is particularly worrisome for QIS applications, since ionizing radiation appears to affect multiple qubits simultaneously. Since quantum error correction (QEC) schemes –necessary for scaling quantum computing– rely on qubits to exhibit random, uncorrelated errors, correlated error sources would render such schemes ineffective. Ionizing radiation has been deemed "catastrophic" because of its ability to potentially circumvent traditional QEC algorithms.

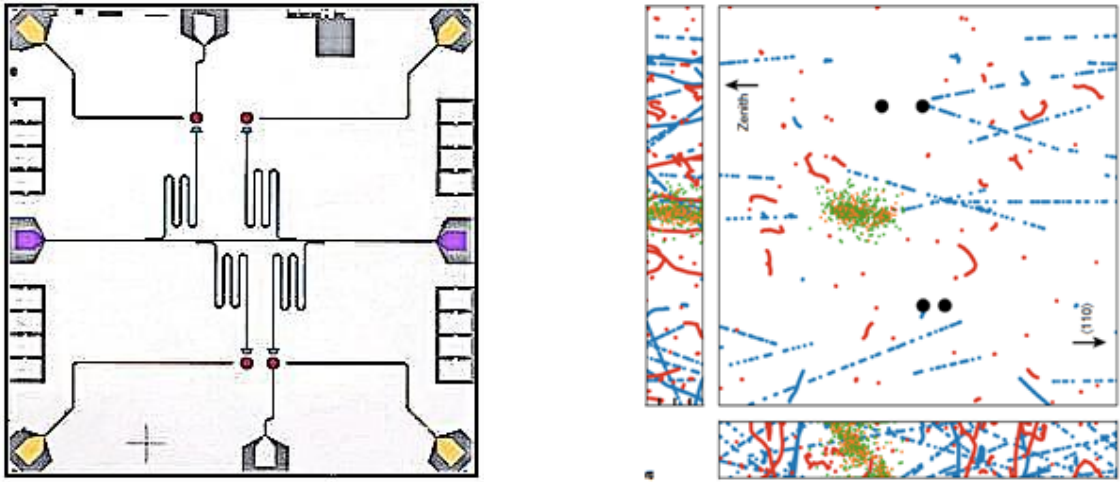

*Figure 1. Layout and simulation of a 4-qubit chip used to investigate the effect of correlated error sources including cosmic rays and background radiation. In the simulation on the right, muons are denoted by blue tracks, photons by red [Nature 594 (2021) 369]. (Image reproduced with permission of the authors.)*

A number of techniques are being devised so as to reduce the impact of radiation. These include spatially distributed error correction schemes, material engineering to promote phonon down-conversion, and developing a more comprehensive understanding of the underlying microscopic physics that leads to quasi-particle poisoning. NP's historic expertise in low radioactivity design, including sourcing and producing materials with low levels of intrinsic radioactivity, can directly help here to reduce the impact of ionizing radiation on qubits. Hybrid devices combining qubits and classical/Quantum 1.0 sensors on a single chip will enable identification of likely quasiparticle poisoning and rejecting calculations in which they occur. This sensor-assisted quantum fault mitigation technique is complementary to quantum error correction (QEC) and, crucially, can identify correlated errors that defeat QEC algorithms. In addition, NP's unique experience in operating sensitive experiments at underground facilities may provide an ideal testbed to test high fidelity qubit systems in low radioactivity environments.

## Sidebar: **Thermal Kinetic Inductance Detectors**

Studies of low energy nuclear physics processes have long played a central role in our efforts to better understand the natural world and future experiments are well positioned to help answer some of the most fundamental of questions: why is there more matter than antimatter in the universe and what are the unseen forces that disappeared as the universe expanded and cooled? The scarcity of new physics at colliders has increased the importance of exploring new and complementary routes to answering these questions. Precision measurements in beta decay are particularly compelling as they can be sensitive to BSM physics at high mass scales, in some cases well beyond that of next generation accelerators. To fully realize this potential, significant increases in precision must be supported through the exploration and application of paradigm shifting technologies.

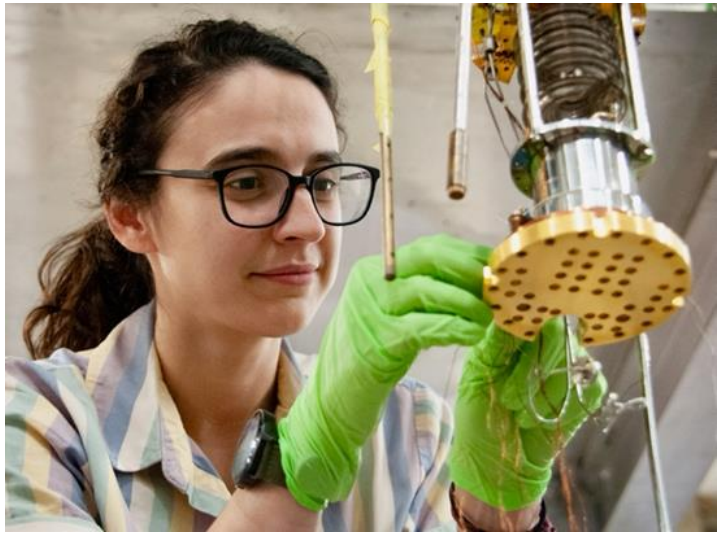

*Figure 2. Dr. Mae Scott developing a low-temperature test-stand for testing of CP-TKIDs with beta sources (Image reproduced with permission from Mae Scott.)*

Particle detection for low energy nuclear physics experiments relies heavily on technologies that are, in essence, decades old and are reaching fundamental limits in performance. Recently, advances in quantum science and technology have benefited precision physics measurements across a variety of fields, from the search for low mass dark matter to background limited astronomy. Detection of particles from nuclear physics processes can also take advantage of these advances. A compelling example is the charged-particle thermal kinetic inductance detector (CP-TKID), a new detection paradigm which combines a conventional thermal calorimeter with a superconducting microwave resonator [Scott2022]. In this type of device, the kinetic inductance of the superconducting resonator is used to accurately measure the change in temperature of a thermally-isolated island due to energy deposited by a charged particle. CP-TKIDs are far more promising than other quantum sensors for scaling to the high-multiplicity large-area detector arrays needed for both full kinematic reconstruction and the low-density sources of some nuclear physics experiments. In addition, calculations suggest that a 10x improvement in energy resolution over existing technology can be achieved, allowing detection of particles with energies well below 1 MeV. Furthermore, previously unexplored CP-TKID topologies may allow for true backscatter suppression supporting precision spectrum and absolute rate measurements.

# Quantum Simulation and Computing

Together with theory and experiment, simulation and computing is a third pillar of Nuclear Science. Classical high-performance computing (HPC) continues to be essential in advancing the NP research program, and NP problems contribute to driving continued expansion of HPC capabilities. The nuclear-theory community has been taking significant steps over the past decade to provide accurate predictions in the realm of many-body NP starting from the underlying *ab initio* theories, according to the roadmap shown in Fig. 1. Unfortunately, the computational resources required to address such problems grow exponentially with the number of quantum-mechanical degrees of freedom in the system (e.g., nucleons in *ab initio* nuclear many-body calculations). Not only that, Monte Carlo importance-sampling techniques suffer from sign problems (or closely-related signal-to-noise degradation) in systems with non-zero fermionic densities and in studies of (real-time) dynamics.

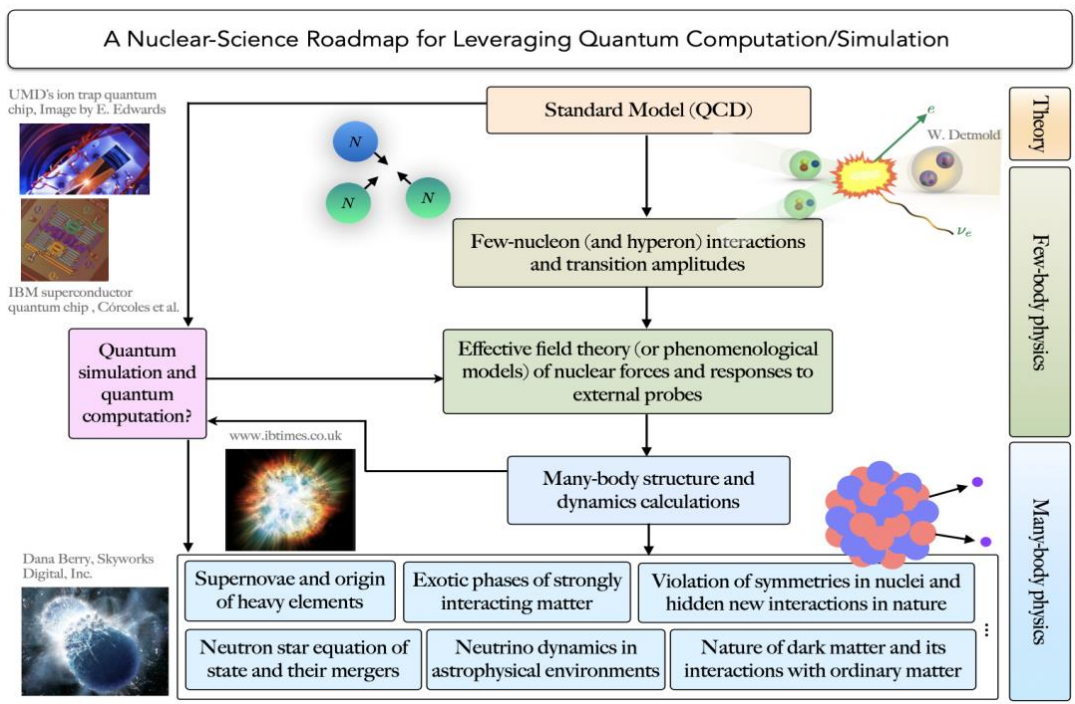

*Figure 3. A roadmap for quantum computing and simulation in NP research. (Image reproduced with permission from Zohreh Davoudi.)*

Therefore, important problems in NP will remain beyond the capabilities of classical HPC. Quantum computing (QC) and quantum simulation (QS) hold the promise for exponential gains to make the required leap in NP many-body problems. Large Hilbert spaces can be encoded exponentially more compactly in (ideal) qubits or other quantum-information-processing units, and quantum superposition and entanglement allow for efficient parallelization of the algorithms that process encoded many-body wavefunctions. In the following, we elaborate on the NP problems that will likely see the largest benefit from QC and QS, how QC and QS are expected to provide such benefits, what the requirements are, and how the advances made in this endeavor may impact other areas of science and technology.

## How will QC and QS advance NP?

*Addressing grand computational problems in NP with QC and QS.* It is expected that QC and QS will uniquely provide predictive capabilities for a range of problems in NP. These include:

- elucidating the phases and phase transitions of strongly-interacting matter governed by QCD, and properties of dense matter in, for example, the interior of neutron stars,
- studying non-equilibrium phenomena such as evolution of matter created in heavy-ion collisions or after the Big Bang, transport phenomena, thermalization dynamics, fragmentation, and hadronization,
- elucidating energy transfer mechanisms in supernovae explosions and neutron-star mergers by simulating coherent neutrino oscillations in dense astrophysical environments,
- constraining electroweak response functions of nucleons and nuclei of relevance to nuclear astrophysics, to searches for violation of fundamental symmetries of the Standard Model, and to the EIC,
- studying low-energy nuclear reactions and fission processes, important for the study of nuclear matter at the limits of stability, and for understanding the formation and role of nuclei in the universe, and more. The pace and form of hardware and algorithmic advances will determine when the community will realistically witness the ultimate quantitative impact of QC and QS in these problems.

*Enhancing classical-computational NP with quantum-information tools.* Ideas and strategies originally rooted in QIST are leading to new developments in classical algorithms. An example is new classical solutions to sign problems informed by QIST principles. In another example, entanglement can be realized as an organizing principle in identifying the most efficient Hamiltonian frameworks and basis considerations in *ab initio* nuclear structure calculations, or in QCD fragmentation algorithms—just as measures of entanglement are used in the development and validation of tensor-network methods in quantum many-body physics.

*Gaining new insights through the lens of entanglement and quantum information.* The drive to cast NP computational problems in the framework of QC and QS in recent years has inspired new efforts in considering entanglement as a probe of nuclear phenomenology, underlying interactions, and emergent symmetries. For example, spin-flavor and other larger 'accidental' symmetries in low-energy nucleon-nucleon and baryon-baryon interactions have been found to be consistent with entanglement suppression. In deep inelastic scattering, parton distribution functions are proposed as a probe of entanglement entropy in the hadrons. In particle colliders, the entanglement among pairs of emitted hadron-antihadrons can probe mechanisms of fragmentation and hadronization after the collision. In non-equilibrium evolution of quantum many-body systems and lattice gauge theories, the entanglement spectrum reveals stages of thermalization, or lack thereof. Last but not least, studies on defining, probing, and harnessing entanglement in quantum fields may lead to deeper insights into the underlying interactions of the Standard Model, and to the design of new distributed quantum-sensing networks in search of new fields and interactions in nature. The continuation of such studies will yield early demonstrations of the transformational impact of QIST in NP even before quantum advantages in QC and QS are obtained.

## What are the strategic goals to accelerate?

*Identifying a roadmap and laying the theoretical and algorithmic groundwork.* The NP community is establishing a roadmap that identifies strategic NP problems suitable for quantum simulation, including an improved understanding of complexity classes of problems of interest, and how physics insights and classical-computational input may be optimally included. The community continues to develop and improve algorithms that translate the NP computational problem to the language of quantum hardware, and is at the early stages of establishing rigorous error bounds informed by properties of the physical system, underlying symmetries, and empirical investigations for each target quantity. Furthermore, efficient Hamiltonian frameworks and state representations consistent with underlying symmetries are being examined, and systematic uncertainties associated with Hilbert-space truncations, simulation-time digitization, and other sources of theoretical and algorithmic error, are being quantified. This groundwork is essential in leveraging both NISQ-era and future fault-tolerant quantum computing systems for NP.

*Leveraging NISQ quantum technologies.* Implementing small instances of NP problems on current NISQ devices is allowing the community to gain quantum-simulation experience, understand the limitations of current and near-term hardware, and improve algorithms. Examples relevant to the quantum simulation of gauge field theories are prototype field theories in lower 1+1 and 2+1 dimensions, and in simpler non-equilibrium quench scenarios that are providing new opportunities for probing non-equilibrium physics of gauge theories, thermalization mechanisms, and the role of entanglement, and may inspire new experimental probes for the high-energy particle-collision experiments. Similarly, low-dimensional models of linear response functions, scattering, and state preparation are forging the path to efficient quantum simulations of nuclear structure and low-energy nuclear dynamics. Another impact of NISQ simulations is in verification, or revealing the shortcomings, of classically-computed predictions in computationally-hard NP problems. For example, NISQ-era simulations are shedding light on the importance of quantum coherence and entanglement in neutrino gasses even in smaller model systems, hence probing mean-field and other semi-classical approximations.

*Technology co-design for NP applications, strengthening cross-disciplinary collaborations, and leveraging quantum testbeds.* The standard paradigm of universal gate-based quantum computing may not be the optimal framework for nearer term advances, given that a broader set of tools in quantum simulators are increasingly available. For example, local multi-dimensional Hilbert spaces and multi-body interactions, the presence of dynamical fermionic and bosonic modes, and the relevance of higher spatial dimensions, are common features required for many NP simulations. Such features can be mapped to similar properties in a range of quantum architectures, including trapped ions, cold atoms in optical lattices, Rydberg arrays controlled by optical tweezers, superconducting circuits in optical

cavities, artificial lattices of dopants in semiconductors, and many more. Conversely, QIST will benefit from co-design opportunities arising from pursuing NP goals, by developing NP-specific quantum-simulation platforms and algorithms, including benchmarking and verification, which may inform future technology developments. For example, custom qubit connectivity and custom gates, modular and distributed systems suited to locality patterns of physical interactions, novel qubits and resonator hardware, and noise-aware error-mitigation techniques can be co-developed and tested for NP applications. Similarly, hardware-aware and problem-informed custom gates can be used in analog simulations, or efficiently interweaved with standard digital gates in hybrid digital-analog quantum computations, to reduce circuit depth and mitigate the impact of decoherence and noise. Such co-design activities can only be pursued through collaborations with technology partners in academia, government sector, and industry, as well as low-level control access to quantum hardware. For example, national DOE testbeds are established to provide an avenue for connecting domain scientists with applications to hardware developers, and nuclear physicists are leveraging these resources to advance near-term NP simulations.

*Expediting development of classical-quantum approaches and leveraging HPC capabilities.* Certain tasks in simulation remain inefficient even for quantum computers and simulators. For example, ground-state preparation for a class of quantum many-body Hamiltonians is known to be QMA-complete. Similarly, preparing thermal (mixed) states is not a natural task in a quantum computer. On the other hand, classical-computational methods in NP have been successful in generating good approximations to ground-state wavefunctions of certain nuclei, or constrain the structure of hadrons and nuclei probed in lattice QCD, which could be used to inform and expedite quantum algorithms for state preparation. Hybrid classical-quantum algorithms such as variational quantum eigensolvers (VQE), that have been advanced for quantum-chemistry applications, have the promise of facilitating spectroscopy and state-preparation tasks in NP. Co-processing algorithms exploiting quantum and classical computers to enact the coordinated propagation of different (e.g., spatial and spin) components of the Hilbert space will play an important role in realizing noise-resilient simulations of nuclear dynamics in the near term. Last but not least, moving further into the NISQ era and beyond, measurement tasks on quantum computers will become increasingly complex, as the number of qubits increases, and may require storing and processing large amounts of data that will only be plausible with HPC resources.

## What do we need in the next 5-10 years?

*Continued, reliable, and timely support of a growing workforce and of activities.* While the prospect of QC and QS for NP was not considered in the last LRP process, the extent of activities and the number of researchers in this area have grown significantly. This is in part due to the rapid advancement of quantum techniques and technologies, increased awareness among domain scientists about the potential of QC and QS in breaking barriers in computationally intractable problems, and increased federal support and new cross-cutting collaborations enabled by the National Quantum Initiative Act. The NP community has produced results that demonstrate how quantum-simulation technologies can be leveraged for a range of problems. However, we are in the early stages of building a robust program of long-term impact. The momentum generated in recent years needs to be maintained, the cross-disciplinary connections and university, government and industry partnerships that have been built need to be sustained, and robust job opportunities should be provided to the growing workforce that is trained in the NP-QIST intersection.

*Broad access to state-of-the-art quantum computers and quantum simulators.* As with classical computing, timely and sustained access to forefront computing hardware and software is key in enabling grand scientific discoveries. NP physicists have been gaining access to quantum-computing devices, from cloud access, to hub access at select universities and national labs, to DOE quantum testbeds, to direct collaborations with individuals at the hardware companies, or with experimentalists with quantum platforms at the universities. As a result, access to forefront quantum computers and simulators has not been uniform among the members of the community. Broad, reliable access to forefront quantum and classical computing systems and their support teams will be essential for advancing NP objectives. Low-level control access to quantum testbeds is particularly desirable to accelerate progress in the co-design of quantum NP algorithms and hardware.

*Active participation in algorithmic and hardware co-design.* As with classical computing, there is likely no single quantum-hardware architecture, and no single universal computing framework, for quantum-simulation developments. It will be a missed opportunity for nuclear scientists to limit their focus to only commercially available qubit and gate-based systems on the market that will likely not be optimized for NP problems. Engagement with hardware developers in academia, government labs, and industry will be the key to expedite problem-specific quantum-computational solutions in NP. This will allow the community to map the problems to the most suitable hardware, and to identify the most suitable computing mode, from digital to hybrid analog-digital to analog settings.

## How does QC and QS in NP impact other NSF and DOE areas?

*Cross-disciplinary connections with High-Energy Physics, Astrophysics, and Cosmology.* Quantum simulation has the potential to provide computationally-feasible approaches to many important questions spanning collider physics, neutrino (astro)physics, cosmology and early-universe physics, and quantum gravity. Simulating systems of relevance to these programs often involve quantum field theories at their core, in the form of Standard Model theories and their effective descriptions, or beyond the Standard Model theories including conformal field theories. Furthermore, accurate NP simulations are essential for experiments that use atomic nuclei for discovering new particles and interactions, and for predicting properties of neutron stars and evolution of dense astrophysical environments. As a result, many of the simulation frameworks developed in NP will have a large overlap with developments in those fields, and many of the NP solutions to be enabled by quantum computing will have a direct impact on High-Energy Physics (HEP), Astrophysics, and Cosmology. As with the classical-computing era, close collaboration among NP and HEP physicists and astrophysicists are anticipated in the quantum-computing era.

*Cross-disciplinary connections with Condensed Matter Physics.* Condensed Matter (CM) studies involve strongly-coupled spin, fermion, and boson systems, often in lower dimensions, as well as phenomena such as (anomalous and chiral) transport, topological phases, superfluidity, and superconductivity. Quantum field theories can emerge as effective low-energy descriptions in such phenomena. For example, topologically ordered states with long range entanglement, relevant for fractional quantum Hall effect and for fault-tolerant topological quantum computation, can be effectively described by lattice gauge theories. Therefore, quantum-simulation tools for gauge theories within NP research can be used in CM research and vice versa. An example of the interplay between the two areas is the development of Tensor Networks as an efficient Hamiltonian simulation method for CM systems with bounded entanglement--a method that was quickly adopted and improved for Hamiltonian simulations of low-dimensional lattice gauge theories in NP and HEP.

*Cross-disciplinary connections with Atomic, Molecular, and Optical Physics.* Parallels between multi-nucleon systems and atomic systems close to unitarity have been exploited by both the NP and Atomic, Molecular, and Optical Physics (AMO) communities, for example, in exploring the role of Efimov states in three-body systems with short-range interactions and in atomic gasses. Unitary atomic fermion systems at large densities pose severe sign problems for classical computing, and development of quantum algorithms for these systems, in NP and AMO, will benefit both communities. Moreover, the AMO community has driven the development of analog and digital quantum-simulation platforms that are being explored in the NP community, and the NP requirements may inform the co-design of next-generation AMO simulation and computing hardware.

*Cross-disciplinary connections with Quantum Chemistry.* Problems in the realm of quantum chemistry and biochemistry are similar to nuclear many-body problems in that they entail describing properties of non-relativistic many-fermion systems, through finding (often approximate) solutions to the corresponding Schrödinger equation. This similarity has led to transfer of many classical-computational tools between the two communities over the years, from Hartree-Fock to post-Hartree-Fock methods, including many-body perturbation theory and coupled-cluster theory. Such cross-disciplinary developments have continued in the quantum-computing era, and will likely grow in the coming years. For example, quantum algorithms for state preparation and spectroscopy using variational quantum eigensolvers with a coupled-cluster ansatz have been commonly explored within both communities.

*Cross-disciplinary connections with Advanced Scientific Computing Research.* The Advanced Scientific Computing Research (ASCR) program within the DOE supports discovering, developing, and deploying computational and networking capability to analyze, model, simulate, and predict complex phenomena of importance to a range of scientific disciplines. It has had a significant role in enabling state-of-the-art HPC in NP through access to DOE's leadership-class facilities, new computing architectures and algorithms, and HPC expertise at National Laboratories. As the ASCR program embarks on developing quantum-computing capability through its targeted programs and initiatives, NP scientists with complex applications, and their growing knowledge and expertise in quantum computing and simulation, will continue to benefit from, and contribute to, ASCR.

*Cross-disciplinary connections with Quantum Information Science.* In the field of Quantum Information Science (QIS), aspects such as topological quantum computing, error correction, and fermion-to-qubit encodings have close connections to gauge theories. For example, fault tolerance of a topological quantum computer arises from a non-local encoding of the states of the quasiparticles that are described by non-Abelian statistics, the Kitaev toric code for error correction realizes a topological phase of matter that exists in $Z_2$ lattice gauge theory, and the local fermionic to qubit maps often use auxiliary degrees of freedom and a set of local (gauge) constraints to enforce Fermi statistics in the qubit description of fermions. As a result, development of simulation frameworks for, and a deeper understanding of entanglement and dynamical properties of, gauge theories can have an impact on both NP and QIS fields alike.

Sidebar: **Quantum Simulation of Neutrino Flavor Evolution in Dense Matter**

The density of neutrinos found in core-collapse supernovae is large enough that they contribute to the transport of energy and momentum, in flavor transport and in dynamics. Coherent evolution of flavors is determined, in part, by electroweak self-interactions among the neutrinos, and the simulation of such systems generally lies beyond the capabilities of classical computation alone. Early studies of the quantum correlations and entanglement in coherent evolution of simple dense neutrino systems, beyond mean-field descriptions, are providing important insights into such dynamics.

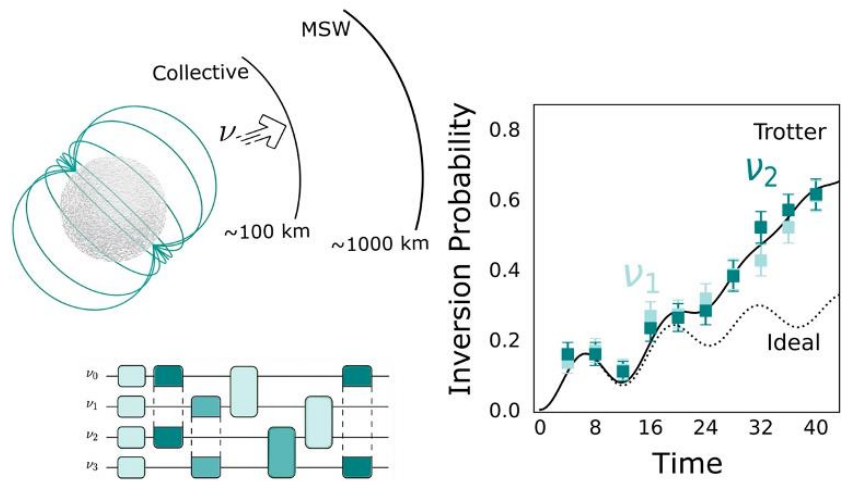

*Figure 4. Quantum simulations of coherent neutrino gasses. The approximate lengths over which coherent neutrino oscillations are relevant during core-collapse supernovae (top left), a cartoon of a quantum circuit implementing time-evolution on a trapped-ion quantum computer (bottom left), and results obtained from a quantum simulation for a specific 4-neutrino system [Phys.Rev.D 107 (2023) 2, 023007] (right). [Image is reproduced with permission from Christian Bauer, Zohreh Davoudi, Natalie Klco, and Martin Savage.]*

Understanding how to simulate systems of neutrinos and quantify their entanglement, which are complex in physical settings, is leading to new insights about their structure and dynamics, such as the recognition of the potential role of dynamical phase transitions, and the potential impact of multi-neutrino entanglement in energy transport and deposition. A number of independent studies and simulations of modest-size systems of neutrinos are pushing the limits of accessible quantum computers, both in the number of qubits and operational fidelity. The neutrino evolution simulated on the quantum devices is currently limited to coherent forward scattering of systems with less than 20 neutrinos, using an effective 2-flavor model, and with an initial tensor-product state

> Sidebar: **Quantum Simulation for Nuclear Dynamics**

First-principles studies of dynamical nuclear many-body phenomena, from nuclear reactions and nuclear response functions to the evolution of matter in the early universe or in particle colliders, is an overarching goal of quantum simulation and computing in NP. While time evolution is expected to be efficiently simulated using ideal quantum computers, states need to first be prepared and then measured to access observables via various strategies.

When it comes to simulations based in the Standard Model, early ground-breaking simulations demonstrated access to real-time phenomena, such as pair production and vacuum fluctuations, in simpler models such as the 1+1 dimensional QED. Algorithms and co-designed protocols for simulating both Abelian and non-Abelian lattice gauge theories in higher dimensions have also been developed, including three-dimensional SU(3) lattice gauge theory. Progress on a range of QCD-inspired problems has been significant and involves algorithms for, and in instances, small hardware implementation of, state preparation, low-lying hadronic spectra, low-energy β-decay amplitude and high-energy particle collisions, parton distributions functions, finite-density thermal phase diagram, transport properties of, and dynamics of quarkonia moving in, quark-gluon plasma, thermalization dynamics and dynamical phase transitions in non-equilibrium states of matter, and fragmentation and hadronization dynamics. While these efforts often involve simple prototype models and lower-dimensional systems, their framework and strategies can be adopted and generalized to QCD.

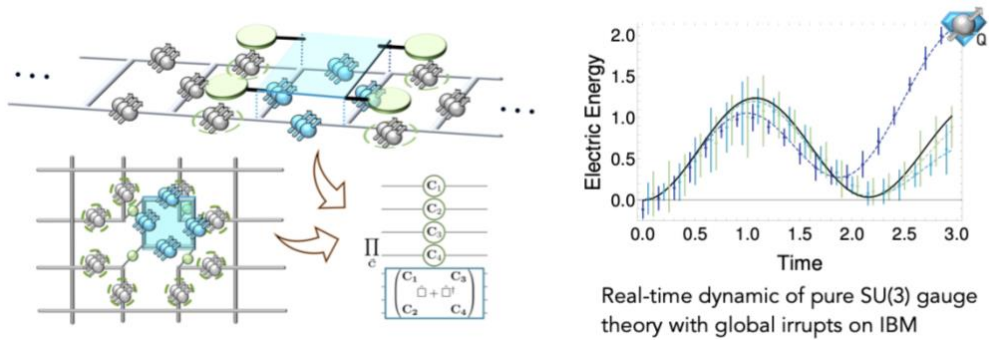

*Figure 5. The structure of the magnetic operator in the pure SU(3) lattice gauge theory upon integration of local quantum numbers for a one-dimensional string of plaquettes (top left) and for a two-dimensional sheet of plaquettes (bottom left). The active quantum registers are denoted by blue squares while the neighboring controls are denoted by green circles. Simulated dynamics of the vacuum in terms of fluctuations in the electric energy for a two-plaquette system using a (truncated) global basis is also shown (right). The simulation was implemented using IBM's Athens quantum processor. The figure has been adopted from [Phys.Rev.D 103 (2021) 9, 094501].*

When it comes to simulations based on nuclear effective Hamiltonians, the first milestone implementations on quantum hardware involved computing the ground state of the deuteron and light nuclei using IBM and Trapped ion devices, followed by subsequent implementations of simple models of radiative processes, and of a nuclear response function. The first quantum algorithm for simulating a generic nuclear Hamiltonian with two-body interactions was developed in the 2000s. This problem did not gain interest until a few years ago, leading to significant growth in algorithmic developments in a range of problems, including neutrino-nucleus scattering and nuclear ground-state preparation. Efforts are also ongoing in designing and benchmarking custom gates and quantum-classical algorithms for nuclear dynamics (see Fig. 4).

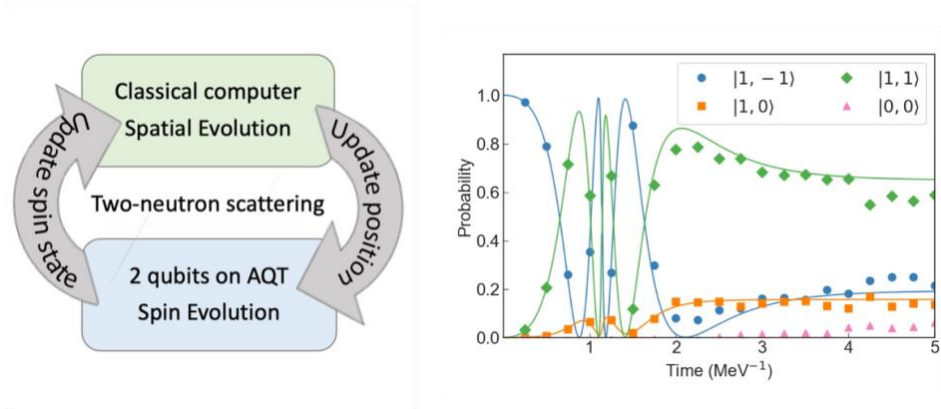

*Figure 6. Schematic representation of the quantum-classical co-processing scheme used to simulate the scattering of two neutrons on the Advanced Quantum Testbed (AQT) at Lawrence Berkeley National Laboratory (left), simulated real time evolution of the spin-state probabilities along the spatial trajectory for the triplet and singlet states of the two-neutron system (right). Solid lines indicate the ideal evolution. Symbols denote the quantum simulation implemented on the AQT. The numbers on the horizontal axis mark arbitrarily chosen stages of the evolution along the trajectory of the relative distance between the two neutrons obtained by solving the classical equation of motion by direct integration. The figure is adopted from [arXiv:2302.06734].*

## Workforce Development

Growing an NP QIST-ready workforce is crucial to advancing NP's objectives. Many of the scientific challenges that we face require an integration of QIST and NP experimental, theoretical, and computational expertise. Since the last LRP, a number of NP scientists have continued to evolve their research focus toward the interface with QIST, and scientists from other areas have changed focus toward QIST-NP problems. The NQI centers have activities focused on NP objectives, involving NP researchers, and are training postdoctoral associates, graduate students, and more senior scientists. Interdisciplinary collaborations born out of smaller Quantum Horizons projects and engagement with DOE-supported quantum testbeds are also training and developing graduate students and postdoctoral associates with a unique blend of skills at the interface of QIST and NP. There are university and national-laboratory activities directed solely at advancing NP-QIST objectives, which include Department of Energy Early Career Awards in the area of QIST-NP, and the InQubator for Quantum Simulation (IQuS). Collaborations with technology companies are advancing NP and providing unique experiences and education, but remain at early stages.

It is desirable that priority be given to continuing to grow and integrate a sustainable quantum-expert workforce at each career stage for NP, to reap the benefits of developments in QIST, and to enable relevant advances in NP to be efficiently utilized to advance QIST. The cross-disciplinary nature of QIST suggests that broad engagement of nuclear physicists with scientists and engineers in other domains is essential to the development of a QIST-ready NP workforce. In addition to growing the workforce, it is essential that viable career paths for researchers at the interface of NP-QIST be sustainable, and recognized as such by early career scientists. A key focus of the proposed *NP Quantum Connection* would be to enhance the fabric of this community of researchers, and especially promoting opportunities across the range of career stages.

It would be beneficial to learn from our experiences in the area of large-scale classical computing and simulation. Forefront computational projects require significant R&D and time investment by computational scientists. These investments include the development of algorithms and software in the co-design process. Such investments should be acknowledged and credited properly.

QIST is a new area in NP research, and due to its multidisciplinary nature, and excitement around its promise, is attracting a more diverse pool of talent into NP. Establishing a *NP Quantum Connection* provides an opportunity to actively and consciously build a diverse and welcoming community, with a researcher population that is representative of society, from the outset.

## Recommendations of the *Quantum Information Science for U.S. Nuclear Physics Long Range Planning* workshop

**Solving grand challenges in Nuclear Physics (NP) requires the development of fundamentally new tools for simulation and sensing. U.S. Nuclear Science is in an early stage of benefiting from and contributing to the advancement of Quantum Information Science and Technology (QIST). To capitalize on this progress, we recommend:**

- **Increasing support for driving advances at the interface of NP and QIST to uniquely address Nuclear Science objectives.**

  This support will advance the development of quantum sensor technology for NP; enhance the (co-)development, integration, and application of quantum-based simulation and computation hardware and techniques for NP; grow cross-cutting research and partnerships that leverage NP expertise to accelerate advances in QIST (including access to forefront hardware and fabrication); and expand the training of, and robust professional pathways for, a diverse and inclusive quantum-ready workforce for NP, with cross-disciplinary collaborations in QIST.

- **Establishing an 'NP *Quantum Connection*' that will realize the transformational potential of QIST in addressing NP grand challenges.**

  This national initiative will enable a community-wide integration of quantum sensing and simulation in NP research; facilitate sharing of resources and expertise among NP, interagency programs, and the national and international QIST community; support bridge junior faculty and scientist positions, postdoctoral fellowships, and graduate and undergraduate students; and strengthen ties with the QIST community, technology companies, and other domain sciences.

# Appendices

## Appendix I: Workshop Participants

**In-person Participants (45)**

Joao Barata, Tanmoy Bhattacharya, Michael Bishof, Ian Cloet, Andrea Delgado, Michael DeMarco, Caleb Fink, Adrien Florio, Marianne Francois, Dorota Grabowska, Shannon Hoogerheide, Mengyao Huang, Kazuki Ikeda, Marc Illa, Kyungseon Joo, Dmitri Kharzeev, Karol Kowalski, Wai Kin Lai, Kyle Leach, Ben Loer, Ian Low, Joshua Martin, David Moore, Thomas Mehen, Niklas Mueller, James Mulligan, Pieter Mumm, Francesco Pederiva, Rob Pisarski, Mateusz Ploskon, Sanjay Reddy, Gautam Rupak, Hersh Singh, Maninder Singh, Ionel Stetcu, Jesse Stryker, Paul Szypryt, Semeon Valgushev, Brent VanDevender, Samuel Watkins, Christopher Wilson and Xiaojun Yao.

**Remote Participants (53)**

Andrei Afanasev, Akif Baha Balantekin, Alessandro Baroni, Raymond Bunker, Bipasha Chakraborty, Ivan Chernyshev, Vincenzo Cirigliano, Benjamin Clark, Shashi Kumar Dhiman, Weijie Du, Dipangkar Dutta, Robert Edwards, Abraham Flores, Alfredo Galindo-Uribarri, Ronald Fernando Garcia Ruiz, Vesselin Gueorguiev, Fanqing Guo, Erin Hansen, Hector Hernandez, Koichi Hattori, Philipp Hauke, Morten Hjorth-Jensen, Keith Jankowski, Calvin Johnson, Denis Lacroix, Dean Lee, Huey-Wen Lin, Xiaohui Liu, Felipe J. Llanes-Estrada, John Looney, Misha Lukin, Alexis Mercenne, Jeff Miller, Emil Mottola, Berndt Mueller, Benjamin Nachman, John Negele, John Orrell, Amol Patwardhan, Daniel Phillips, Stephen Poole, Irene Qualters, Mike Rumore, Thomas Schaefer, Jeremy Scott, Rajeev Singh, James Vary, Juan-Jose Galvez-Viruet, Kyle Wendt, Hongxi Xing, Liang Yang, Glenn Young, Fanyi Zhao

**Workshop Co-organizers (6)**

Douglas Beck, Joseph Carlson, Zohreh Davoudi, Joseph Formaggio, Sofia Quaglioni and Martin Savage.

## Appendix II: Workshop Presentations

**Overviews**

- *Quantum simulation for Nuclear Physics* by Sofia Quaglioni (LLNL) and Zohreh Davoudi (UMD)
- *Superconducting sensors and related technology* by Pieter Mumm (NIST)
- *Quantum sensors in Nuclear Physics and future directions* by Doug Beck (UIUC)
- *Programmable quantum systems for simulations and sensing* by Misha Lukin (Harvard)

**Quantum Sensing**

- *Low background cryogenic facility @ PNNL* by Ben Loer (PNNL)
- *Searches for massive neutral particles emitted in nuclear decays with mechanical quantum sensors* by David Moore (Yale)
- *Rare isotope doped superconducting tunnel junctions as a precision tool for BSM physics searches* by Kyle Leach (Colorado School Mines)

**Quantum Field Theories and QCD**

- *Quantum simulations of real-time QCD processes* by Dmitri Kharzeev (Stony Brook)
- *Entropy suppression through quantum interference in electric pulses* by Adrien Florio (BNL)
- *Simulating bosonic theories with bosonic quantum circuits* by Michael DeMarco (BNL/MIT)
- *Jet evolution in nuclear matter* by Joao Barata (BNL)
- *"QZD": a toy model for QCD* by Rob Pisarski (BNL)
- *Towards calculating first-principles strong interactions on universal quantum computers* by Jesse Stryker (UMD)
- *Heavy-ion collisions - emergent phenomena in QCD* by Mateusz Ploskon (LBNL)
- *Entanglement suppression and emergent symmetries in low-energy QCD* by Ian Low (Northwestern)
- *Efficient quantum simulations of lattice gauge theories* by Dorota Grabowska (IQuS/UW)
- *Quantum simulation of jet quenching in nuclear environments produced in heavy-ion and electron-ion collisions* by Xiaojun Yao (IQuS/UW)

**Dense Neutrinos in Astrophysical Environments**

- *Entanglement in three-flavor collective oscillations* by Baha Balantekin (Wisconsin)
- *Multi-body entanglement in dense neutrino systems* by Marc Illa (IQuS/UW/QSC)
- *Neutrino flavor oscillations* by Josh Martin (LANL)

**Analog Quantum Simulators**

- *Analog quantum simulation of topological models with a parametric cavity* by Christopher Wilson (IQC/Waterloo)
- *Towards quantum simulations of Nuclear Physics Using Yb optical tweezer arrays* by Michael Bishof (ANL)

**Quantum Many-Body Theory and Nuclear Physics**

- *Toward hybrid (classical/quantum) computing for many-body problems* by Karol Kowalski (PNNL)
- *Nuclear Physics and QIS: challenges and opportunities* by Niklas Mueller (IQuS/UW)
- *Trapped ion quantum simulation of collective neutrino oscillations* by Francesco Pederiva (Trento)
- *Quantum machine learning applications to Nuclear Physics* by Andrea Delgado (ORNL)
- *Inelastic reactions on a quantum computer* by Gautam Rupak (Mississippi State)
- *Noise resilient quantum simulations for Nuclear Science* by Kyle Wendt (LLNL)
- *State preparation with a projection algorithm* by Ionel Stetcu (LANL)
- *Transitions in (1+1) light front $\varphi^4$ theory using quantum computing method* by Mengyao Huang (Iowa State)

# Appendix III: Endorsement of support for NP-QIST in other LRP events

The mutual role, and the need for supporting the growth, of QIST and NP has been acknowledged and endorsed during the National Science Advisory Committee's 2023 Long-Range Planning process, including in two community Town Hall meetings on i) Nuclear Structure, Reactions, and Astrophysics, and on ii) Fundamental Symmetries, Neutrons, and Neutrinos, sponsored by the Division of Nuclear Physics (DNP) of the American Physical Society (APS), as well as an LRP workshop on Computational Nuclear Physics and AI/ML. The excerpts of the relevant text of the resolutions, recommendations, and/or endorsements are provided in this appendix for reference.

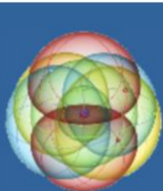

**NSAC Long Range Plan Town Hall Meeting on Nuclear Structure, Reactions and Astrophysics**

Nov 14 – 16, 2022
Argonne National Laboratory
US/Central timezone

## Resolution 3

Computing is essential to advance all fields of nuclear science. We strongly support enhancing opportunities in computational nuclear science to accelerate discoveries and maintain U.S. leadership by:

- Strengthening programs and partnerships to ensure the efficient utilization of new high-performance computing (HPC) hardware and new capabilities and approaches offered by artificial intelligence/machine learning (AI/ML) and quantum computing (QC);

- Establishing programs that support the education, training of, and professional pathways for a diverse and multidisciplinary workforce with cross-disciplinary collaborations in HPC, AI/ML, and QC;

- Expanding access to dedicated hardware and resources for HPC and new emerging computational technologies, as well as capacity computing essential for many research efforts.

# Fundamental Symmetries, Neutrons, and Neutrinos Town Meeting

13-15 December 2022
UNC - Chapel Hill
America/New_York timezone

Fundamental Symmetries, Neutrons, and Neutrinos Town Meeting participants endorse the following recommendations by its Quantum Information Science subcommittee:

Nuclear physics has much to gain from, and contribute to, quantum information science (QIS) and quantum sensing. Following the recommendations of an NSAC Subcommittee on QIS in 2019:

**We recommend increased investment to capitalize on the rapid worldwide development of quantum sensor technology and its timely implementation in NP.**

**We recommend investment in exploratory research directions that aim to develop, integrate, and apply quantum-based simulation and computation techniques in NP.**

**We recommend strengthening the QIS expertise in NP training to create a diverse, quantum-ready, nuclear-capable workforce.**

**We recommend further development of research that leverages the knowledge base of NP that can help inform and solve outstanding problems in QIS, and to encourage cross-cutting research between the two research communities.**

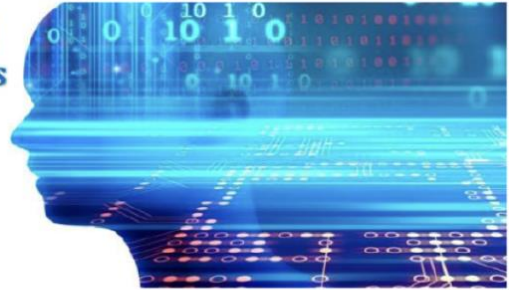

Computational Nuclear Physics and AI/ML Workshop

## Workshop Resolution

High-performance computing is essential to advance nuclear physics on the experimental and theory frontiers. Increased investments in computational nuclear physics will facilitate discoveries and capitalize on previous progress. Thus, we recommend a targeted program to ensure the utilization of ever-evolving HPC hardware via software and algorithmic development, which includes taking advantage of novel capabilities offered by AI/ML.

The key elements of this program are to:

1) Strengthen and expand programs and partnerships to support immediate needs in HPC and AI/ML, and also to target development of emerging technologies, such as quantum computing, and other opportunities.

2) Take full advantage of exciting possibilities offered by new hardware and software and AI/ML within the nuclear physics community through educational and training activities.

3) Establish programs to support cutting-edge developments of a multi-disciplinary workforce and cross-disciplinary collaborations in high-performance computing and AI/ML.

4) Expand access to computational hardware through dedicated and high-performance computing resources.